%% file: _main.tex
\documentclass[sigconf]{aamas}

\usepackage{balance} 
\input{math_commands.tex}

\setcopyright{ifaamas}
\acmConference[AAMAS '25]{Proc.\@ of the 24th International Conference
on Autonomous Agents and Multiagent Systems (AAMAS 2025)}{May 19 -- 23, 2025}
{Detroit, Michigan, USA}{A.~El~Fallah~Seghrouchni, Y.~Vorobeychik, S.~Das, A.~Nowe (eds.)}
\copyrightyear{2025}
\acmYear{2025}
\acmDOI{}
\acmPrice{}
\acmISBN{}

\acmSubmissionID{630}

\title[AAMAS-2025 Formatting Instructions]{\algo: Task-Agnostic Contrastive pre-Training for \\ Inter-Agent Communication}

\author{Peihong Yu}
\affiliation{
  \institution{University of Maryland}
  \city{College Park}
  \country{USA}}
\email{peihong@umd.edu}

\author{Manav Mishra}
\affiliation{
  \institution{IISER Bhopal}
  \city{Bhopal}
  \country{India}}
\email{mishra20@iiserb.ac.in}

\author{Syed Zaidi}
\affiliation{
  \institution{University of Maryland}
  \city{College Park}
  \country{USA}}
\email{szaidi@terpmail.umd.edu}

\author{Pratap Tokekar}
\affiliation{
  \institution{University of Maryland}
  \city{College Park}
  \country{USA}}
\email{tokekar@umd.edu}

\input{AAMAS2025/sections/abstract}


\newcommand{\BibTeX}{\rm B\kern-.05em{\sc i\kern-.025em b}\kern-.08em\TeX}

\newcommand{\algo}{\textsc{TACTIC}}

\begin{document}


\pagestyle{fancy}
\fancyhead{}


\maketitle 

\input{AAMAS2025/sections/introduction}
\input{AAMAS2025/sections/relatedWork}
\input{AAMAS2025/sections/preliminary}

\input{AAMAS2025/sections/method}
\input{AAMAS2025/sections/experiments}
\input{AAMAS2025/sections/conclusion}




\begin{acks}
If you wish to include any acknowledgments in your paper (e.g., to 
people or funding agencies), please do so using the `\texttt{acks}' 
environment. Note that the text of your acknowledgments will be omitted
if you compile your document with the `\texttt{anonymous}' option.
\end{acks}



\bibliographystyle{ACM-Reference-Format} 
\bibliography{AAMAS2025/sample}


\end{document}

%% file: math_commands.tex
\usepackage{graphicx}
\usepackage{subcaption}
\usepackage{multirow}
\usepackage{bm}
\usepackage{placeins}
\usepackage{afterpage}
\usepackage{enumerate}

\usepackage{amsmath,amsfonts,bm}

\def\eqref#1{equation~\ref{#1}}

\def\1{\bm{1}}

\def\va{{\bm{a}}}

\DeclareMathAlphabet{\mathsfit}{\encodingdefault}{\sfdefault}{m}{sl}
\SetMathAlphabet{\mathsfit}{bold}{\encodingdefault}{\sfdefault}{bx}{n}

\def\gA{{\mathcal{A}}}

\def\gC{{\mathcal{C}}}

\def\gN{{\mathcal{N}}}

\def\gS{{\mathcal{S}}}
\def\gT{{\mathcal{T}}}



%% file: AAMAS2025/sections/abstract.tex
\begin{abstract}
The ``sight range dilemma'' in cooperative Multi-Agent Reinforcement Learning (MARL) presents a significant challenge: limited observability hinders team coordination, while extensive sight ranges lead to distracted attention and reduced performance. While communication can potentially address this issue, existing methods often struggle to generalize across different sight ranges, limiting their effectiveness. We propose \algo, Task-Agnostic Contrastive pre-Training strategy Inter-Agent Communication. \algo{} is an adaptive communication mechanism that enhances agent coordination even when the sight range during execution is vastly different from that during training. The communication mechanism  encodes messages and integrates them with local observations, generating representations grounded in the global state using contrastive learning. By learning to generate and interpret messages that capture important information about the whole environment, \algo\ enables agents to effectively ``see'' more through communication, regardless of their sight ranges. We comprehensively evaluate \algo\ on the SMACv2 benchmark across various scenarios with broad sight ranges. The results demonstrate that \algo\ consistently outperforms traditional state-of-the-art MARL techniques with and without communication, in terms of generalizing to sight ranges different from those seen in training, particularly in cases of extremely limited or extensive observability.
\end{abstract}


%% file: AAMAS2025/sections/introduction.tex
\section{Introduction}





\begin{figure}
    \centering
    \includegraphics[width=0.85\linewidth]{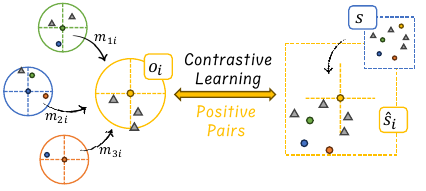}
    \caption{\algo\ utilizes contrastive learning to align the integration of local observations $o_i$ and messages $\{m_{ji}\}$ with the full egocentric state $\hat{s}_i$ for each agent $i$, enabling agents to "see" beyond their limited sight ranges through communication.}
    \label{fig:overview}
\end{figure}


Multi-agent Reinforcement Learning (MARL) provides a framework for addressing complex coordination tasks across various domains such as robotics~\cite{zhang2021multi,yang2004multiagent,busoniu2008comprehensive}, autonomous vehicles~\cite{qu2024modelassisted,shalev2016safe}, and network optimization~\cite{zhang2019integrating,li2022applications}. In MARL, agents often operate under partial observability, where each agent's perception is limited to a certain ``sight range'' around itself, which results in a fundamental challenge known as the \textit{sight range dilemma}~\cite{shao2023complementary}. 
The dilemma lies in balancing tension between an agent's need for local information to make decisions and the broader context required for effective team coordination. Agents with narrow sight ranges often struggle to coordinate effectively due to limited environmental information, while those with extensive sight ranges can become overwhelmed by excessive data, leading to inefficient learning and reduced performance. 

\begin{figure}
    \centering
    \includegraphics[width=0.49\linewidth]{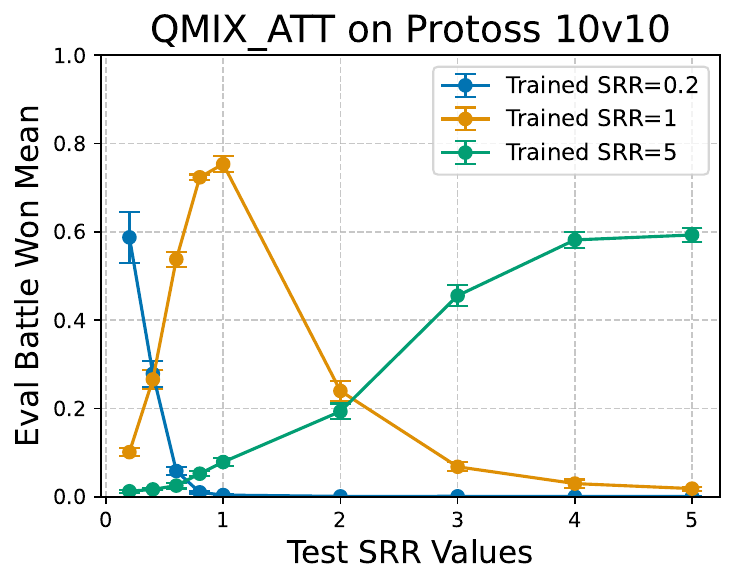}
    \includegraphics[width=0.49\linewidth]{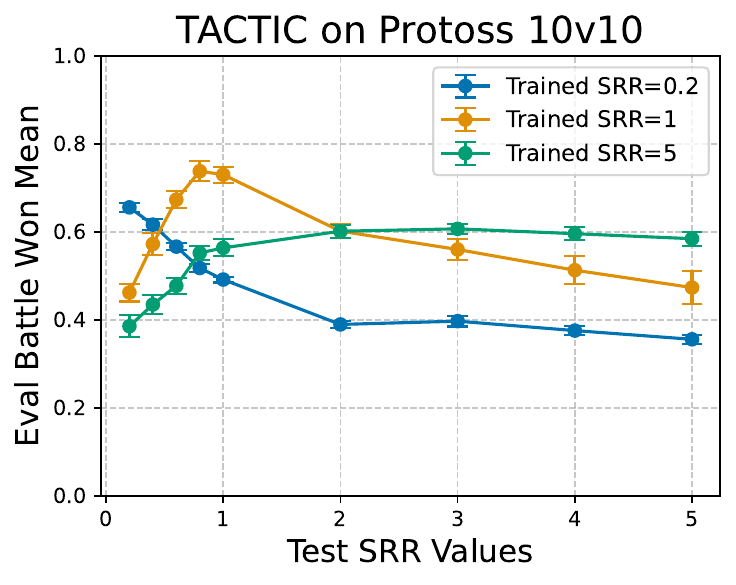}
    \caption{QMIX-ATT and \algo's performances on Protoss 10v10 from SMACv2~\cite{ellis2024smacv2} with varying sight ranges~(SRs). Different SRs are achieved by applying different sight-range ratios~(SRRs) to the agents' original SRs in the implementation. Policies trained at SRR=0.2, 1, and 5 are tested across a broader set of SRRs. QMIX-ATT Struggles to generalize to unseen SRs, while \algo\ generalizes much better.}
    \label{fig:motivating}
\end{figure}

Researchers have proposed various approaches to addressing this challenge, primarily focusing on communication mechanisms that allow agents to share information. 
Those methods include targeted communication strategies~\cite{das2019tarmac, sukhbaatar, liu2020who2com}, attention mechanisms~\cite{niu2021multi, seraj2021heterogeneous, she2022agent}, or graph-based methods~\cite{shen2021graphcomm, sun2020scaling, hu2024learning}. 
For example, QMIX-Att \cite{yang2020qattengeneralframeworkcooperative} integrates attention into the QMIX framework for selective communication, and TarMAC \cite{das2019tarmac} uses signature-message pairs for context-aware communication. While effective, these methods assume the same sight ranges during training and during execution, limiting their capacity to adapt to varying visibility conditions, as illustrated in Figure \ref{fig:motivating}.

Generalizing across varying sight ranges provides considerable benefits for MARL systems. It facilitates more efficient and cost-effective deployments by enabling a single, adaptable model to handle diverse observability conditions, eliminating the need for separate models for each scenario. This flexibility is crucial in real-world applications, where systems must adjust to different visibility conditions. For example, autonomous vehicles need to adapt to visibility fluctuations due to weather or time of day, while search and rescue robots may encounter visual obstructions from debris or smoke. 
In this work, we are particularly interested in the cases where the sight range during execution is fixed but different from that seen during training.

We hypothesize that \textit{agents can better generalize across different sight ranges if they can communicate in a way that leads to a more comprehensive understanding of the global environment}. Based on this hypothesis, we propose a novel approach that aligns the integrated local observations and messages with each agent's egocentric (global) state, as illustrated in Figure \ref{fig:overview}.
The egocentric state serves as an ideal alignment target, providing a comprehensive yet agent-specific view of the environment during training. We use contrastive learning to achieve this alignment, encouraging agents to develop a communication protocol that bridges the gap between limited local observations and the broader environmental context.
This process enables agents to effectively ``see'' more through communication, regardless of their actual sight ranges. 

In addition to the contrastive learning objective, we introduce two auxiliary losses: a reconstruction loss and a dynamics loss. The reconstruction loss helps ensure that the learned representations retain essential information from the original observations, while the dynamics loss encourages the model to capture the temporal relationships in the environment. Crucially, our method is task-agnostic in nature, as it does not rely on task-specific reward information when learning to communicate, solely focusing on capturing the underlying environment information, which further enhances the flexibility and adaptability of our method across diverse scenarios. 




To this end, we introduce \textbf{\algo} (\textbf{T}ask-\textbf{A}gnostic \textbf{C}ontrastive pre-\textbf{T}raining for \textbf{I}nter-Agent \textbf{C}ommunication), a novel strategy designed to enhance generalization across varying sight ranges in cooperative MARL.
TACTIC operates through two key stages: 
(1) \textbf{Offline contrastive pretraining}, where we use contrastive learning on an offline dataset to pretrain two key communication modules: a message generator and a message-observation integrator.
(2) \textbf{Online policy integration}, where the pre-trained communication modules are frozen and incorporated into agents' online policy learning, enabling dynamic communication adaptation during task execution while preserving the learned task-agnostic properties.

We summarize the main contributions of this work as follows:
\begin{itemize}
    \item A task-agnostic communication mechanism that enables adaptive message generation and interpretation;
    \item A cooperative MARL framework with communication called \algo\ that alleviates the sight range dilemma;
    \item A comprehensive evaluation of \algo\ in the SMACv2 environment showing \algo's superior performance regarding generalizability across sight ranges and training efficiency.
\end{itemize}
Our experimental results on the SMACv2 (StarCraft Multi-Agent Challenge) benchmark show that TACTIC outperforms existing state-of-the-art MARL with communication techniques. Our method demonstrates robust generalization capabilities, enabling effective coordination across various sight ranges (Figure~\ref{fig:motivating}).

%% file: AAMAS2025/sections/relatedWork.tex
\section{Related Work}

\textbf{Communication in MARL.} In cooperative MARL, communication is a critical component for addressing partial observability and improving agent coordination ~\citep{sukhbaatar2016learning, lo2024learning}. Learning effective communication in this context involves several challenges, including determining \textit{who} communicates, \textit{how} messages are conveyed, and \textit{what} information is transmitted under bandwidth or sight-range constraints. 

Several approaches have been proposed to address these challenges. Targeted communication methods focus on identifying specific agents for message exchange, while graph-based and attention-based models structure communication based on relational dynamics between agents. Graph-based methods, such as the graph-attention network proposed by Niu et al.~\cite{niu2021multi}, enable agents to dynamically adjust communication based on relevance, improving scalability in complex environments. Techniques that manage bandwidth limitations address the need to optimize when and what information should be shared. For example, attention-based methods such as TarMAC~\cite{das2019tarmac} leverage signature-message pairs and attention mechanisms to enable dynamic, context-aware communication between agents. Information-theoretic approaches, such as NDQ~\cite{wang2019learning}, introduce regularization to minimize communication overhead while maximizing the informativeness of messages.

Recent advancements also aim to make multi-agent communication more interpretable and flexible. Lin et al.~\cite{lin2021learning} proposed grounding communication by autoencoding raw observations into messages, allowing agents to develop a shared understanding of communication symbols. Similarly, MASIA~\cite{guan2022efficient} aggregates raw observations into latent representations that can be used to reconstruct the global state, providing agents with a more holistic view of the environment. Du et al.~\cite{du2021learning} introduced methods to learn correlated communication topologies, which reduce redundancy and optimize coordination among agents by refining communication pathways. In parallel, Zhang et al.\cite{zhang2021succinct} introduced Temporal Message Control (TMC), a technique that applies temporal smoothing to reduce the number of inter-agent messages, achieving robust and efficient communication in resource-constrained environments without sacrificing performance.


Our work builds on these advancements by addressing adaptive communication under sight-range limitations, proposing a novel approach that optimizes communication frequency and content based on changing agent observations. This differs from previous works by integrating bandwidth and perceptual constraints into a unified framework, enabling efficient communication without reliance on pre-defined structures.

\textbf{Contrastive Learning in MARL.} Contrastive learning is a representation learning technique that aims to bring similar (positive) samples closer together in the learned feature space while pushing dissimilar (negative) samples farther apart. This is typically achieved using a contrastive loss function.. Methods such as Contrastive Predictive Coding (CPC) by Oord et al. \cite{oord2018representation} laid the groundwork for learning predictive representations by contrasting positive and negative samples. Building on this, TACO \cite{taco} adapts contrastive learning to RL by learning useful representations through temporal abstraction. More recently, the use of supervised contrastive loss \cite{khosla2020supervised}, which incorporates multiple positive and negative samples per anchor point, has been explored. This extension enables richer representation learning by capturing a broader set of relevant relationships, which is particularly valuable in RL tasks with multiple favorable outcomes.

In the multi-agent domain, contrastive learning has seen adaptation for both non-communicative and communicative settings. For MARL without communication, methods like COLA \cite{cola} have demonstrated the ability to improve coordination among agents by using contrastive objectives to refine agent policies based on shared goals. This approach emphasizes the utility of contrastive learning in situations where direct agent-to-agent communication is absent or limited. In contrast, for settings where agent communication is possible, contrastive learning has been utilized to optimize communication strategies between agents. Lo et al. \cite{lo2024learning} apply contrastive learning to improve multi-agent communication, enabling agents to develop more efficient communication protocols that reduce unnecessary message exchanges while maintaining performance. Additionally, methods such as the one proposed by Singh et al. \cite{singh2019learning} focus on learning when to communicate, which is crucial in reducing communication overhead in resource-constrained environments. Zhang et al. \cite{zhang2019efficient} further refine this by incorporating variance-based control mechanisms, allowing agents to communicate only when necessary, improving overall system efficiency.

Our work extends these ideas by integrating contrastive learning into multi-agent communication under sight range limitations. This approach differs from previous work in that we train the communication module offline specifically with the goal of generalizing across sight ranges.

%% file: AAMAS2025/sections/preliminary.tex
\section{Preliminaries}
\textbf{Dec-POMDP with Communication.} We consider the fully cooperative MARL problem with communication, which can be modeled as Decentralized Partially Observable Markov Decision Process (Dec-POMDP) \cite{oliehoek2016concise} and formulated as a tuple $\langle \gN, \gS, \gA, P, \Omega, O, R, \gamma, \gC \rangle$. The sets $\gN=\{1,...,n\}$ denotes the indexing of the agents, $\gS$ is the state space, $\gA$ is the action space, $\Omega$ is observation space, and $\gC$ denotes all possible communication messages. Each agent $i\in\gN$ acquires an observation $o_i=O(s, i)\in \Omega$, where $O$ is the observation function and $s\in\gS$. A joint action $\va = \langle a_1, ..., a_n \rangle$ leads to the next state $s' \sim P(s'|s, \va)$ and a shared global reward $r = R(s,\va)$ where $R$ is the reward function.  

Each agent selects actions based on the observation-action history $\tau_i \in \gT \equiv (\Omega \times \gA)^{*}$~\footnote{$^*$ denotes the product over time} using a policy $\pi(a_i|\tau_i, m_i)$ where $m_i = [m_{ji} \in \gC, j \in \gN]$ denotes the incoming messages for agent $i$ and $m_{ji}$ is the message sent from agent $j$ to agent $i$. The policy is shared across agents during training. 

The overall objective is to find a joint policy $\bm{\pi}(\bm{\tau}, \va)$ to maximize the global value function 
\begin{equation}
Q^{\boldsymbol{\pi}}(\boldsymbol{\tau}, \boldsymbol{a})=\mathbb{E}_{s, \boldsymbol{a}}\left[\sum_{t=0}^{\infty} \gamma^t R(s, \boldsymbol{a}) \mid s_0=s, \boldsymbol{a}_{\mathbf{0}}=\boldsymbol{a}, \boldsymbol{\pi}\right],
\end{equation}
where $\bm{\tau}$ is the joint observation-action history of all agents and $\gamma \in [0,1)$ is the discount factor. We follow the \textit{Centralized Training and Decentralized Execution} (CTDE) paradigm and adopt the architecture of QMIX~\cite{rashid2018qmix} to form our algorithm~\footnote{For clarity, we drop the time superscripts for states and actions}.


\noindent \textbf{Contrastive Learning.}
Contrastive learning is a powerful paradigm in representation learning, particularly in the context of deep learning, where it aims to learn embeddings by contrasting positive and negative samples. The fundamental idea is to pull together representations of similar instances (positives) while pushing apart those of dissimilar instances (negatives). In the supervised setting, the SupCon (Supervised Contrastive) loss \cite{khosla2020supervised} extends this framework by allowing for multiple positive samples per anchor, thereby leveraging label information more effectively. The loss objective can be mathematically expressed as:
\begin{equation}
 L_{supcon} = \sum_{i} \frac{-1}{|P(i)|} \sum_{p \in P(i)} \log \left( \frac{\exp(z_i \cdot z_p / \upsilon)}{\sum_{a \in A(i)} \exp(z_i \cdot z_a / \upsilon)} \right)    
\end{equation}
where $z_i$ denotes the normalized embedding of the anchor sample, $P(i)$ is the set of positive samples corresponding to the anchor, $A(i)$ is the set of all samples in the batch excluding the anchor, and $\upsilon$ is a temperature parameter that controls the sharpness of the distribution. This formulation enhances the clustering of similar instances in the embedding space and improves robustness against natural corruptions.

%% file: AAMAS2025/sections/method.tex
\begin{figure*}[htbp]
    \centering
    \includegraphics[width=0.95\linewidth]{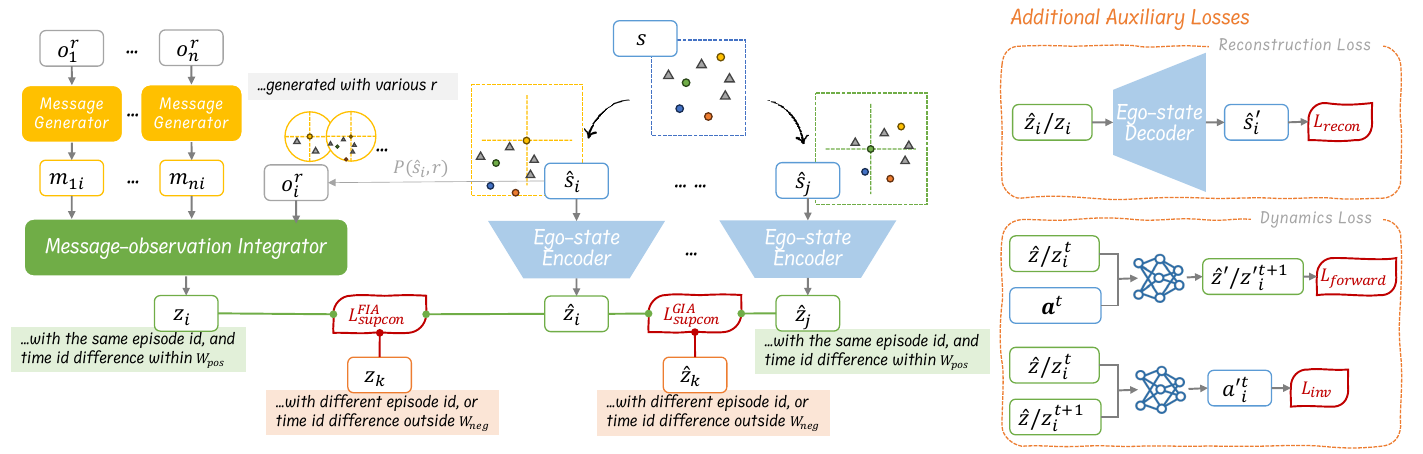}
    \caption{
    The offline training pipeline of the adaptive communication mechanism(Section~\ref{sec:offlineTrain}). It includes three key components: an egocentric state encoder, an adaptive message generator, and a message-observation integrator. The training pipeline consists of two contrastive learning processes: Global Information Alignment (GIA) for aligning the features generated from the egocentric state encoder across all agents and timesteps, and Feature Integration Alignment~(FIA) for aligning features from the message-observation integrator and the egocentric state encoder on an individual agent. 
    Two auxiliary loss functions are introduced in the total loss function to enhance training: a deconstruction loss for learning to recover the egocentric states and a dynamic loss for learning temporally coherent representations.
    }
    \label{fig:OffLineTraining}
\end{figure*}

\section{\algo: Towards task-agnostic adaptive communication in MARL}

In this section, we introduce our algorithm \algo.
\algo\ consists of two stages: (1) an offline training of a communication mechanism that works well with varied sight ranges, and (2) an online training of the agent coordination policy.
The task-agnostic offline training stage solely utilizes the environment states and local observations of agents and doesn't rely on environmental reward signals or any policy.
The online policy training stage learns a task-specific policy with the communication module from the first stage frozen.
We explain the offline stage in Section~\ref{sec:offlineTrain} and the online stage in Section~\ref{sec:onlineTrain} in detail.

\subsection{Offline Training of Communication Mechanism}
\label{sec:offlineTrain}

One main challenge in Multi-Agent Reinforcement Learning (MARL) is to develop effective joint policies when each agent has only a partial observation of the environment. By learning to communicate more effectively, agents are expected to overcome their limited observability and achieve better coordination. Furthermore, the mechanism used to generate and use messages between agents should be flexible enough to handle different sight ranges.

To this end, we present an approach where, in an offline training stage, we use contrastive learning on a pre-collected dataset to develop a communication mechanism that can adapt to different observation ranges. The rationale is that, by learning to create and integrate messages that capture important information about the whole environment, agents can effectively \textbf{``see'' more through communication}, regardless of their current sight range.

The offline dataset $\mathcal{D}$ consists of a set of trajectories, where each trajectory $\tau = \{(s^t, o_{1:n}^t, a_{1:n}^t)\}_{t=1}^T$ represents a sequence of $T$ timesteps for $n$ agents. The dataset can be collected through random exploration of the environment~(details in Section~\ref{sec:ablation}).
%
The overall offline training pipeline for the adaptive communication mechanism is illustrated in Figure~\ref{fig:OffLineTraining}, where three key components are present: an egocentric state encoder, an adaptive message generator, and a message-observation integrator:  
\begin{enumerate}
    \item The \textit{egocentric state encoder} takes an egocentric state $\hat{s}_i$ and generates its corresponding feature embedding $\hat{z}_i$, where the $\hat{s}_i$ are obtained from the global state $s$, which preserves all information from $s$ but represents from the perspective of agent $i$.
    \item The \textit{adaptive message generator} takes the partial observation $o_i^r$ of an agent with \textit{varying} sight ranges and outputs the message $\{m_{ij}\}_{j=1}^n$ it communicates to other agents. We obtain $o_i^r$ by randomly sampling a sight range $r$ and applying a masking operation over the egocentric state $\hat{s}_i$, denoted as $P(\hat{s}_i, r)$.
    \item The \textit{message-observation integrator} takes an agent's partial observation $o_i^r$ and all the messages $\{m_{ji}\}_{i=1}^n$ it receives from other agents, integrating them into a feature embedding $z_i$.
\end{enumerate}

With the three components at hand, the offline training consists of two contrastive learning processes:
\begin{enumerate}
    \item \textit{Global Information Alignment~(GIA)}: a supervised contrastive (SupCon) loss for aligning the feature embeddings $\hat{z}_i$ generated from the egocentric state encoder across all agents, ensuring that the egocentric state encoder captures consistent and relevant information from agents about the whole environment;
    \item \textit{Feature Integration Alignment~(FIA)}: a supervised contrastive (SupCon) loss for aligning the integrated feature $z_i$ calculated by the message-observation integrator with the generated feature $\hat{z}_i$ from the egocentric state encoder for each specific agent, pushing the adaptive message generator to learn to synthesize the most informative messages to communicate and allowing the message-observation integrator to reflect a more complete picture of the environment~(i.e., the agent’s egocentric state) given the agent's limited observation and the messages they receive.
\end{enumerate}
In FIA, the function $P(\hat{s}_i, r)$ applies augmentation by randomly varying the sight range $r$ (resulting in $o_i^r$). This process exposes the adaptive message generator to diverse scenarios with changing sight ranges, rather than fixed ones, thereby enhancing its generalizability.
For contrastive learning in both GIA and FIA, we define positive and negative pairs based on the offline dataset $\mathcal{D}$. Features from agents within the same episode (i.e., from the same trajectory) and within a timestep window of length $W_{pos}$ form positive pairs. Conversely, features from agents in different episodes or separated by more than $W_{neg}$ timesteps constitute negative pairs. Note that both GIA and FIA are task-agnostic as they don't interact with any environmental reward signals.


To further improve the learned representations, we incorporate two additional auxiliary learning objectives: (1) a reconstruction loss~($L_{recon}$): a decoder network learns to recover the egocentric state $\hat{s}_i$ from the feature embeddings produced by either the egocentric state encoder or the message-observation integrator; and 
(2) a dynamic loss~($L_{dyn}$): this includes both forward and inverse dynamics predictions using MLP networks. The forward model predicts $\hat{z}_i^{t+1}$ (or $z_i^{t+1}$) 
given $\hat{z}_i^t$ (or $z_i^t$) and $\boldsymbol{a}^t$, while the inverse model predicts ${a}_i^t$ given consecutive feature embeddings $(\hat{z}_i^{t},\ \hat{z}_i^{t+1})$ or $(z_i^{t},\ z_i^{t+1})$. These auxiliary objectives promote comprehensive, temporally coherent representations that better support adaptive communication in multi-agent scenarios.



The final loss for the overall offline training is a weighted sum of the contrastive losses and the auxiliary losses:
\begin{equation}\label{eq:loss}
L_{total} = L_{supcon}^{GIA} + L_{supcon}^{FIA} + \alpha L_{recon} + \beta L_{dyn}    
\end{equation}
where $\alpha$ and $\beta$ are weighting factors.

\subsection{Online Training of Agent Policy}
\label{sec:onlineTrain}

\begin{figure}
    \centering
    \includegraphics[width=0.9\linewidth]{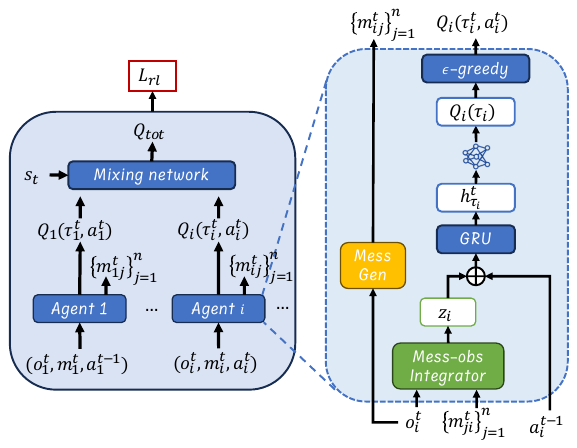}
    \caption{Online policy training pipeline of \algo, illustrating the integration of QMIX architecture with pre-trained communication components. The pre-trained message generator (Mess Gen) and message-observation integrator (Mess-obs Integrator) remain fixed during the policy training.}
    \label{fig:online}
\end{figure}

After offline pretraining, we integrate the \textit{adaptive message generator} and \textit{message-observation integrator} into the QMIX \cite{rashid2018qmix} framework for online policy training~(illustrated in Figure \ref{fig:online}).  
In this stage, each agent $i$ processes its observation and receives messages~(synthesized with the \textit{adaptive message generator} from other agents) to generate an integrated representation $z_i$ using the \textit{message-observation integrator}. This representation $z_i$, together with the agent's last action, informs the agent's action selection via a GRU-based Q-network. 
The QMIX architecture then combines individual agents' Q-values through a mixing network conditioned on the global state to produce a centralized Q-value, which is used to compute the loss for training agent policies and the mixing network. Importantly, the parameters of the pre-trained \textit{adaptive message generator} and \textit{message-observation integrator} remain fixed during this online training phase and are not updated.

In the remaining sections, we present the experimental setup and results that evaluate the effectiveness of our methodology.


%% file: AAMAS2025/sections/experiments.tex
\section{Experiments}

\begin{figure*}
    \centering
    \includegraphics[width=\linewidth]{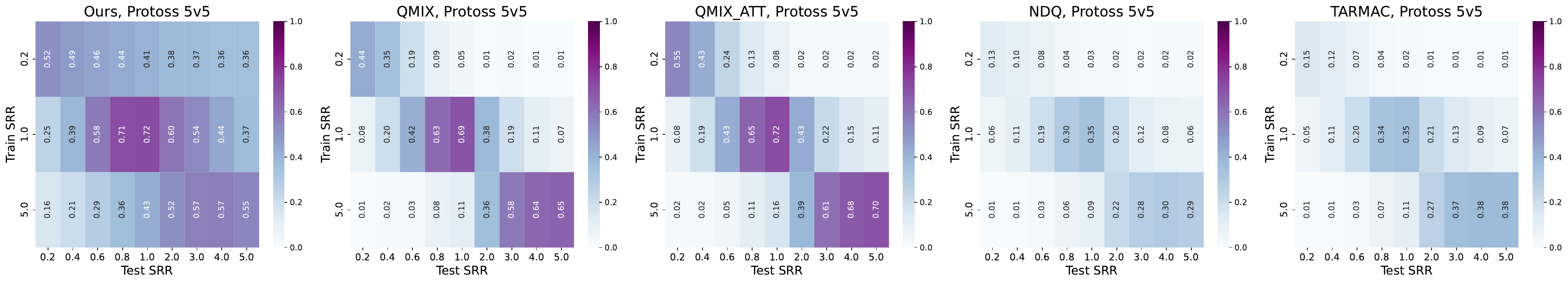}
    \includegraphics[width=\linewidth]{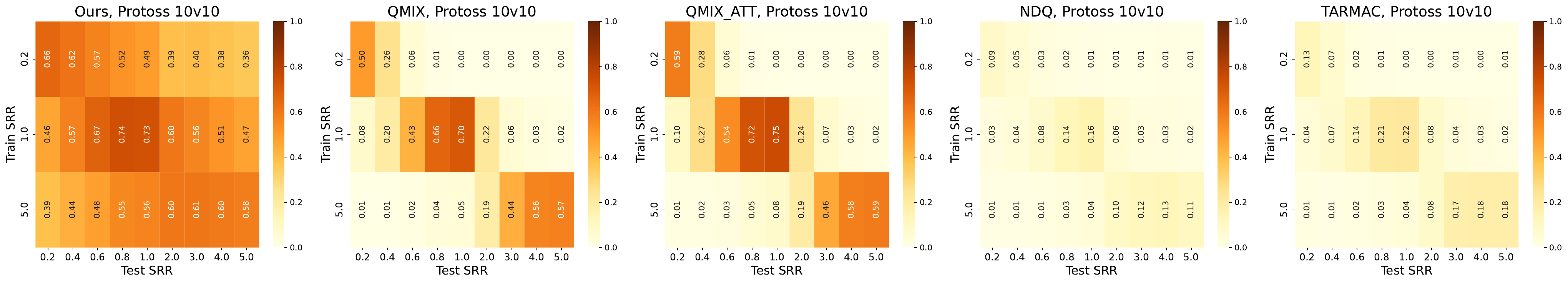}
    \includegraphics[width=\linewidth]{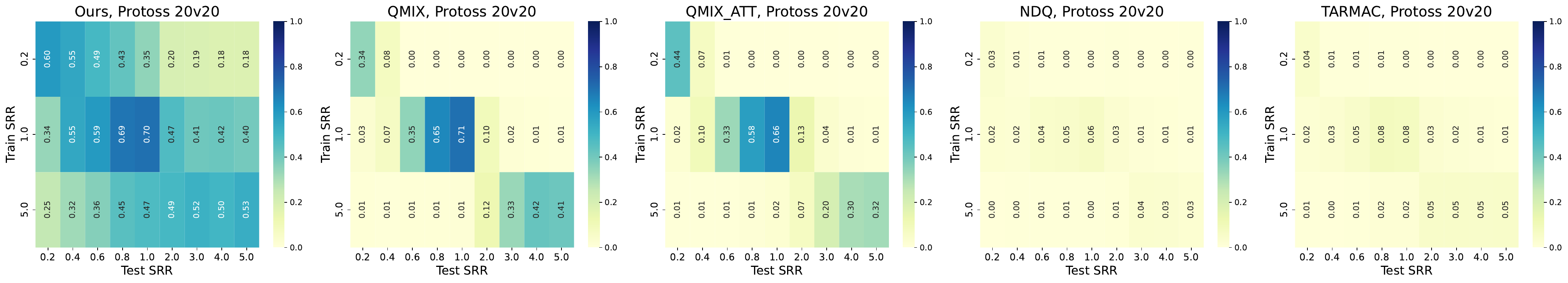}
    \caption{Performance of \algo\ and baseline models on policy generalizability across various sight ranges in the Protoss map.}
    \label{fig:generalization-protoss}
\end{figure*}

\begin{figure*}
    \centering
    \includegraphics[width=\linewidth]{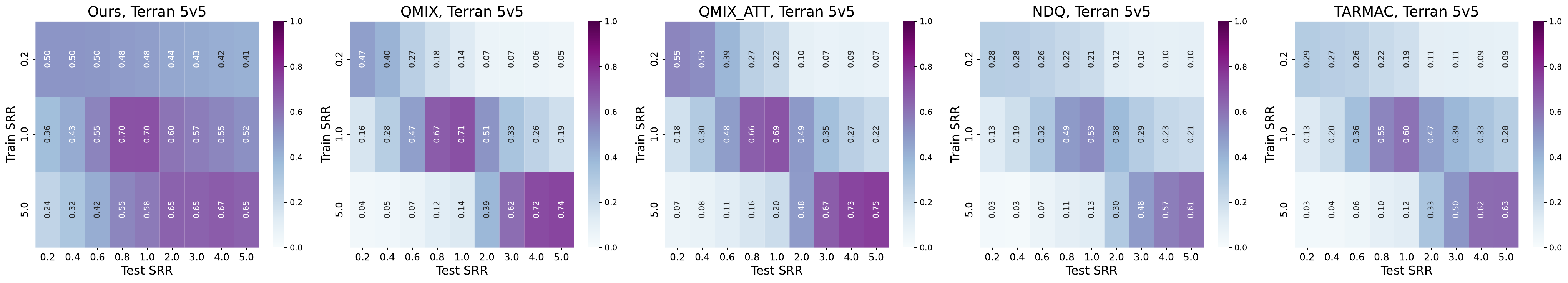}
    \includegraphics[width=\linewidth]{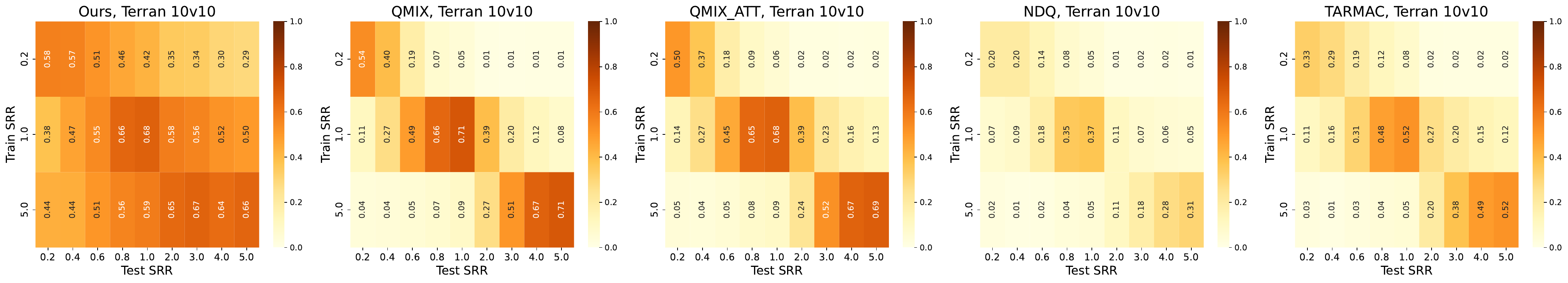}
    \includegraphics[width=\linewidth]{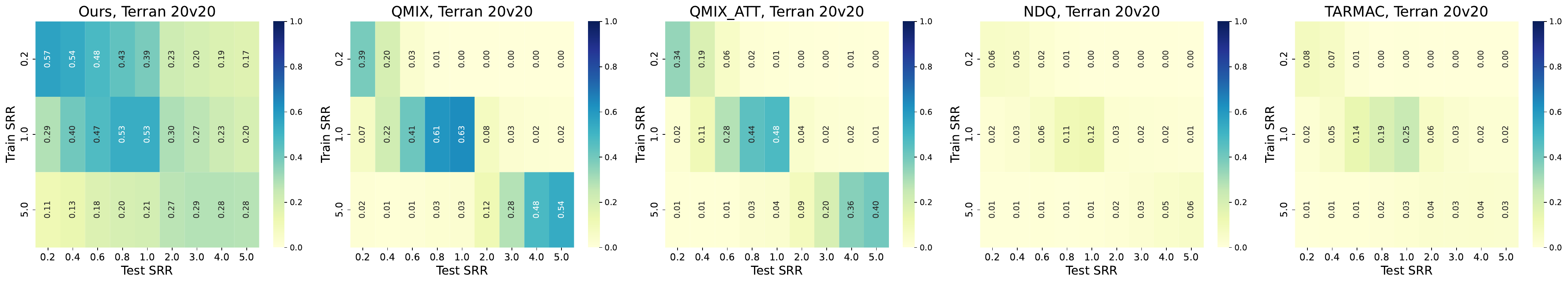}
    \caption{Performance of \algo\ and baseline models on policy generalizability across various sight ranges in the Terran map.}
    \label{fig:generalization-terran}
\end{figure*}

\begin{figure*}
    \centering
    \includegraphics[width=\linewidth]{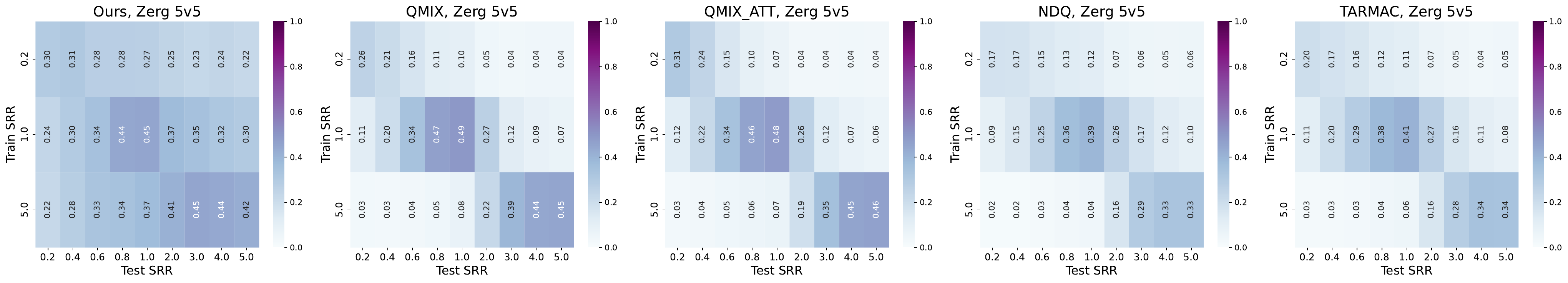}
    \includegraphics[width=\linewidth]{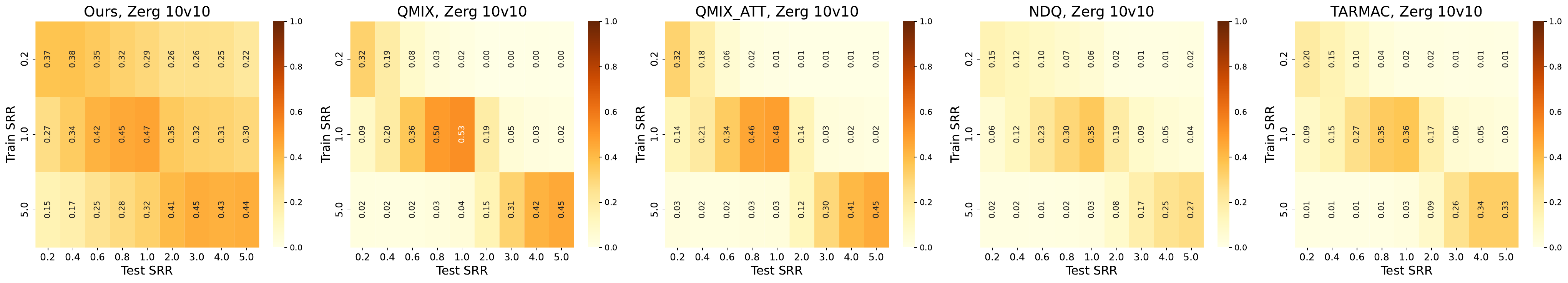}
    \includegraphics[width=\linewidth]{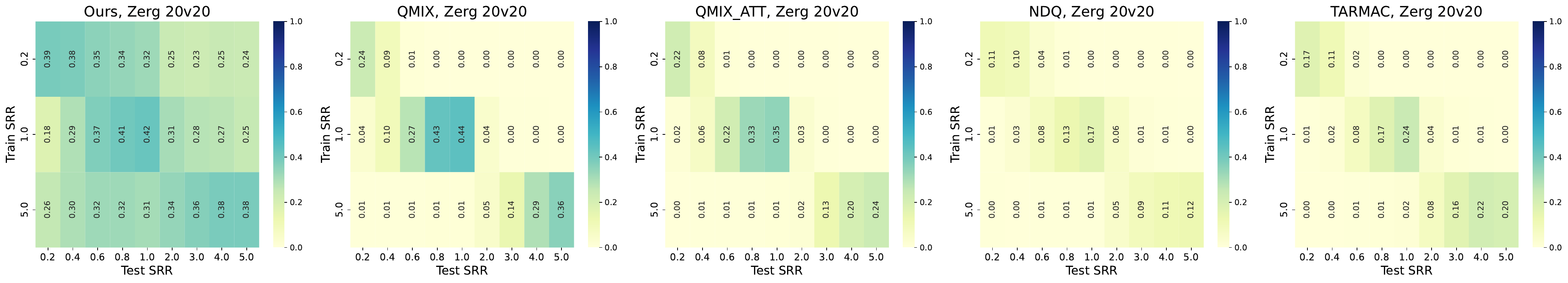}
    \caption{Performance of \algo\ and baseline models on policy generalizability across various sight ranges in the Zerg map.}
    \label{fig:generalization-zerg}
\end{figure*}

In this section, we report our evaluation of \algo.
Our experiments aim to address the following key questions:

\begin{enumerate}
    \item[Q1.] Can the policy trained with the adaptive communication mechanism generalize across different sight ranges?
    \item[Q2.] Does the offline-trained communication mechanism enhance the efficiency of online policy training?
    \item[Q3.] How do data quality and varying loss terms impact the performance of training effectiveness and generalization?
\end{enumerate}

\subsection{Experimental Setup}\label{sec:expSetup}
Our experiments are conducted in the SMACv2 environment~\cite{ellis2024smacv2}. Compared to the original SMAC environment~\cite{samvelyan2019starcraft}, SMACv2 incorporates increased stochasticity and meaningful partial observability, necessitating the development of complex closed-loop policies for effective agent coordination. We use three maps from SMACv2 for our experiments: Terran, Protoss, and Zerg. Each map features distinct unit types, with Terran units including Marines, Marauders, and Medivacs; Protoss units comprising Stalkers, Zealots, and Colossi; and Zerg units consisting of Zerglings, Hydralisks, and Banelings. Agents are generated procedurally, with varying numbers ranging from 5 to 20, and are assigned specific sight and attack ranges that enhance the complexity of the scenarios. 


For each map, we consider three agent number configurations (5, 10, and 20 agents) and three sight-range ratios (0.2, 1, and 5). The sight-range ratios are applied by multiplying the agents' original sight ranges, which vary among different agents, to adjust their visibility accordingly. This allows us to assess the adaptability of the proposed communication mechanism under varying observability conditions.



For each environment setup specified by a combination of the map and the agent number configuration, we pre-train an offline communication mechanism~(Section~\ref{sec:offlineTrain}).
This communication mechanism is then used for online policy training~(Section~\ref{sec:onlineTrain}) with a team of agents with fixed sight ranges. The learned policy is further evaluated on a variety of sight ranges~(the sight range ratios are between 0.2 and 5) on the same environment setup.

To valid the effectiveness of \algo, the performance regarding the generalization across various sight ranges and the online training efficiency is compared to four baselines: QMIX~\cite{rashid2018qmix}, QMIX-Att~\cite{yang2020qattengeneralframeworkcooperative}, NDQ~\cite{wang2019learning}, and TarMAC~\cite{das2019tarmac}. QMIX is one of the more commonly used CTDE MARL algorithms that does not use communication during execution. The others are three variants of QMIX with different communication mechanisms. We adopted the hyperparameters from the original implementations of the baselines and reused the hyperparameters of QMIX for \algo. All results in the following sections are from five independent runs with different random seeds.

\subsection{Policy Generalization Across Sight Ranges}
\label{sec:exp_generalization}
To answer Q1, in this section, we report the generalization capability of our trained policies across various sight ranges. As introduced in Section~\ref{sec:expSetup}, our policies are learned based on a pre-trained communication mechanism under the same environment setup with fixed sight ranges; they are then tested on a broader set of sight-range ratios valued from 0.2 to 5. Their performances~(i.e., \textit{mean battle won rate}) in all scenarios are recorded and visualized with heatmaps as shown in Figure~\ref{fig:generalization-protoss}, \ref{fig:generalization-terran} and \ref{fig:generalization-zerg}.

It can be clearly seen that our method \algo\ produces policies that demonstrate more robust performances in all the environment settings and across various sight-range ratios that were not seen during policy training.
The four baselines can have reasonable performances only when the test and train sight ranges are close~(referring to the diagonal cells of the heatmaps for the baselines), while policies learned by \algo\ are capable of maintaining satisfying performances even when the train and test sight ranges are different by large. For example, on the map Terran with 5 agents, policy trained with sight-range ratio $0.2$ by \algo\ has a $0.41$ \textit{mean battle won rate} when tested with sight-range ratio $5$, while the corresponding results for the four baselines are all under $0.10$; it is also true if exchanging the train and test sight-range ratios or in other maps. With more agents introduced into the environment, baseline models become more strict on the difference between train and test sight ranges~(i.e., their heatmaps become more sparse with more zero values), while our method \algo\ still generalizes well.

The observed phenomenon suggests that our adaptive communication mechanism that is learned offline helps agents better generate and interpret messages and develop flexible strategies to take actions to adapt to changing observability conditions, leading to \textbf{reusable policies} in environments with unseen sight ranges. However, for the four baseline approaches, one needs to \textit{re-train} the model to get a usable policy if the observation conditions change.


\subsection{Online Policy Training Efficiency}\label{sec:ablation}
\begin{figure*}
    \centering
    \includegraphics[width=0.7\linewidth]{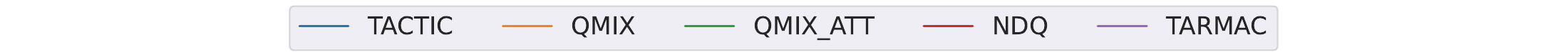}
    \includegraphics[width=0.49\linewidth]{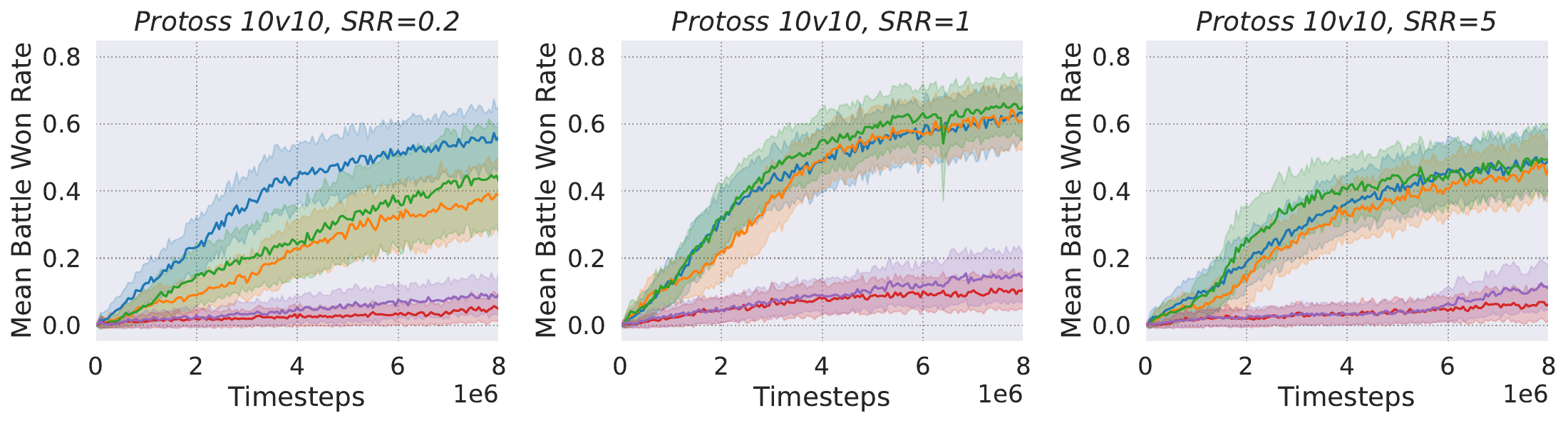}
    \includegraphics[width=0.49\linewidth]{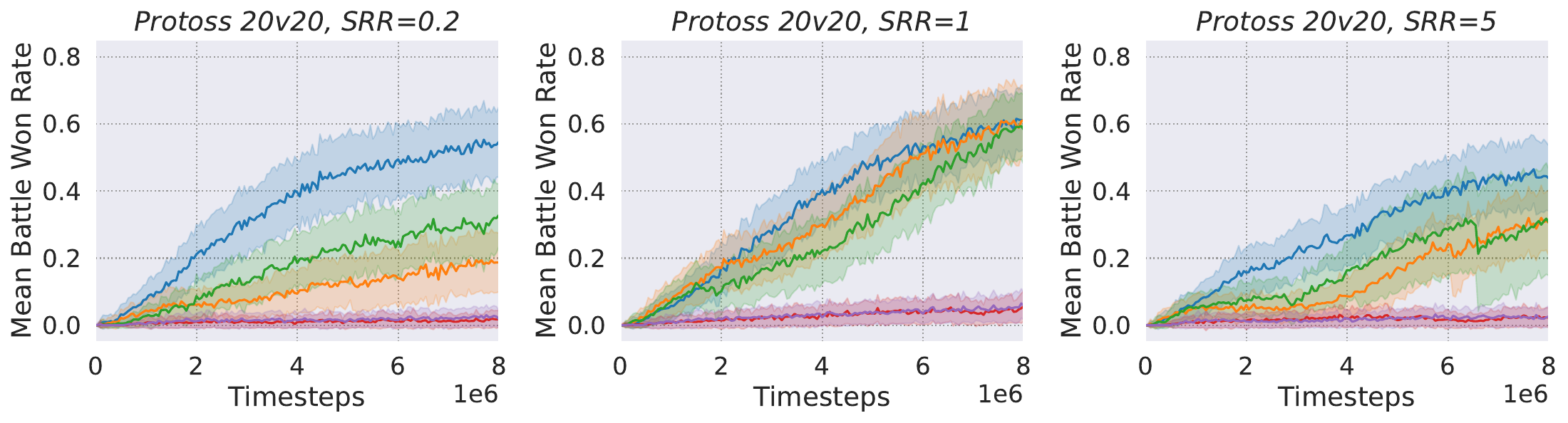}
    \includegraphics[width=0.49\linewidth]{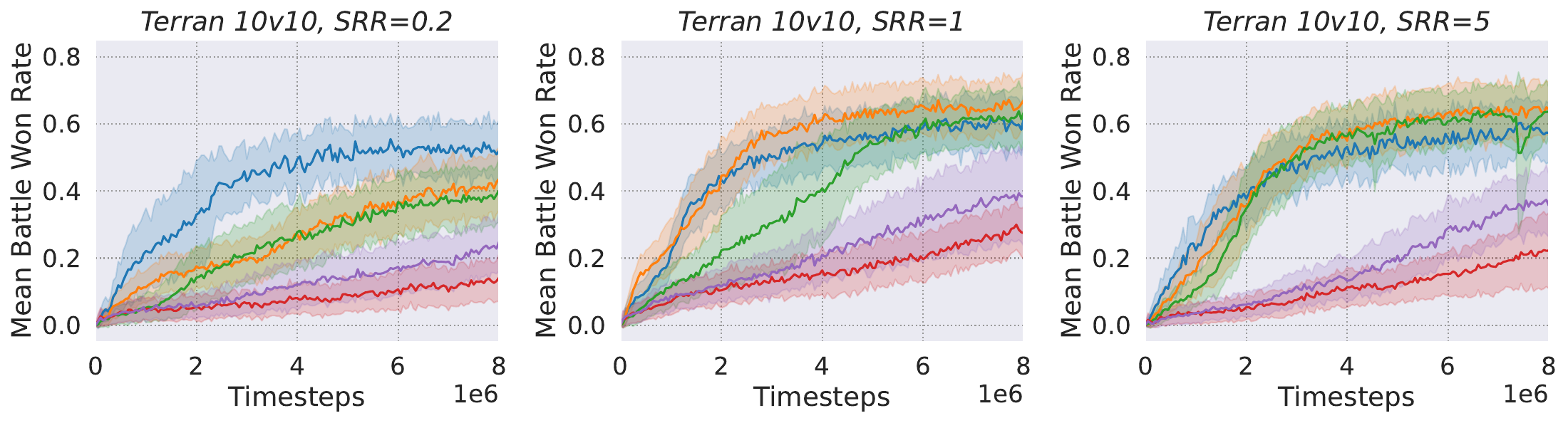}
    \includegraphics[width=0.49\linewidth]{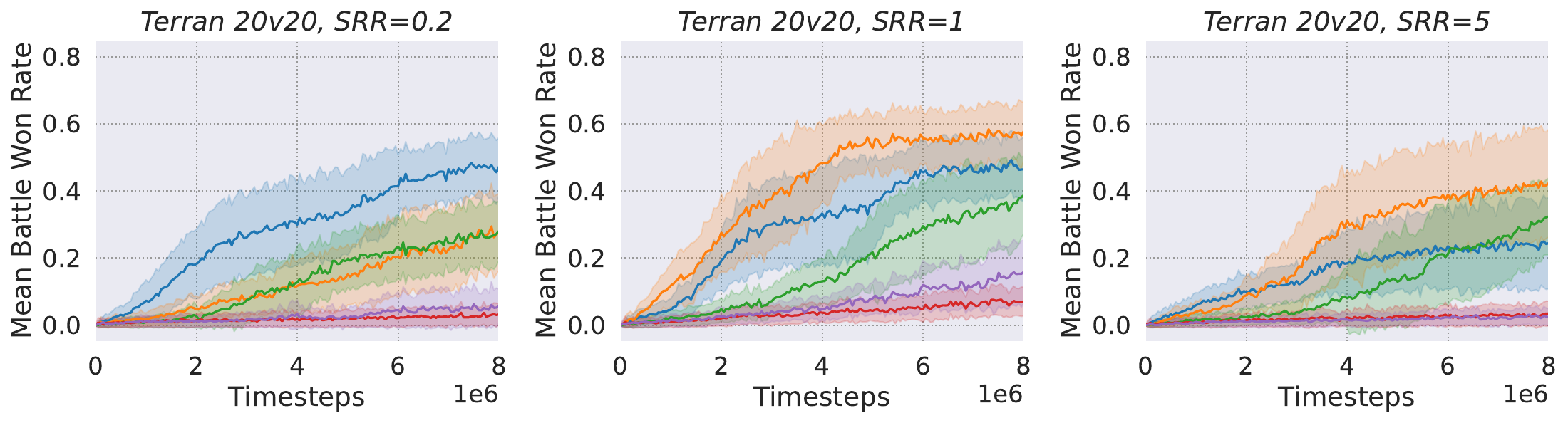}
    \includegraphics[width=0.49\linewidth]{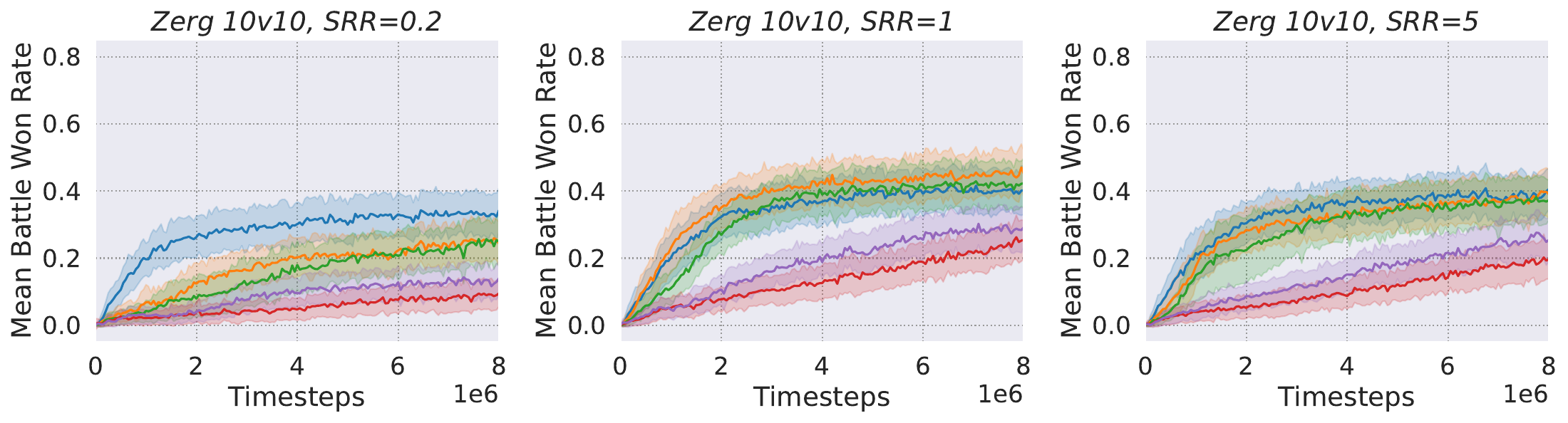}
    \includegraphics[width=0.49\linewidth]{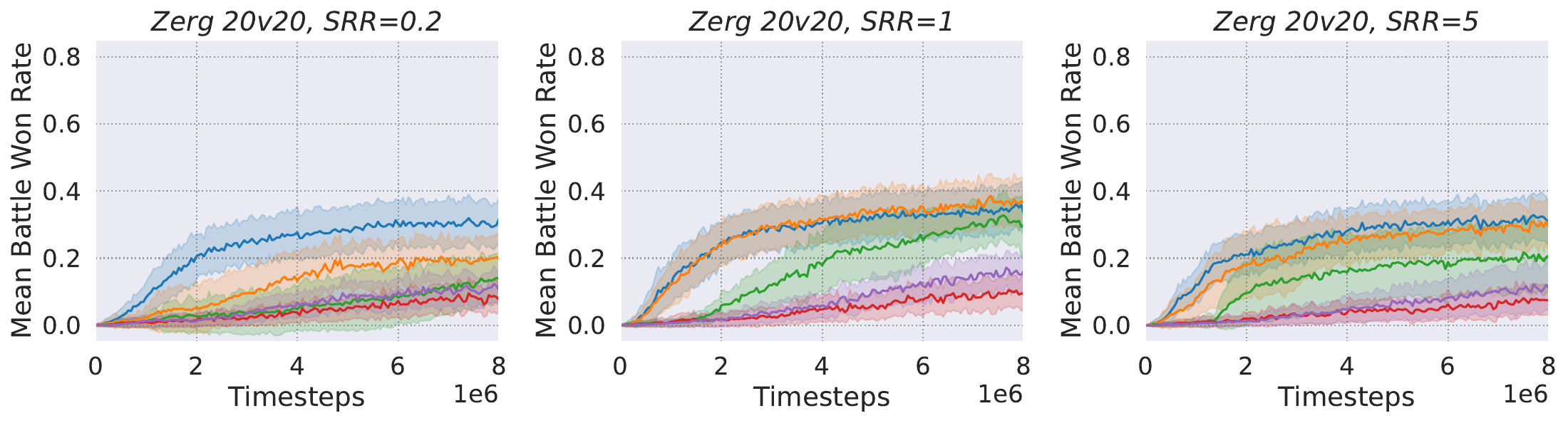}
    \caption{Training curves of the online policy learning stage in \algo\ and baseline models under different environment setups.}
    \label{fig:training_efficiency}
\end{figure*}

To answer Q2, we compare the online policy training efficiency of our method \algo\ against the four baselines. Figure~\ref{fig:training_efficiency} presents the learning curves~(\textit{mean battle won rate} versus \textit{timesteps}) of all five algorithms on every combination of map type, number of agents, and the sight-range ratio. 

It can be observed that in every environment setup, the convergence speed of \algo\ is superior to or at the same level as that of QMIX or QMIX-Att, and is always better than that of NDQ and TarMAC. Notably, when the sight-range ratio used during online policy training is small~(i.e., $0.2$), \algo\ demonstrates significantly better convergence speed compared to the baselines. 

The results indicate that the offline-trained communication mechanism contributes positively to the training efficiency of online policy learning, especially in scenarios where the number of agents is large or the sight ranges of agents are small.


\begin{figure*}
    \centering
    \includegraphics[width=\linewidth]{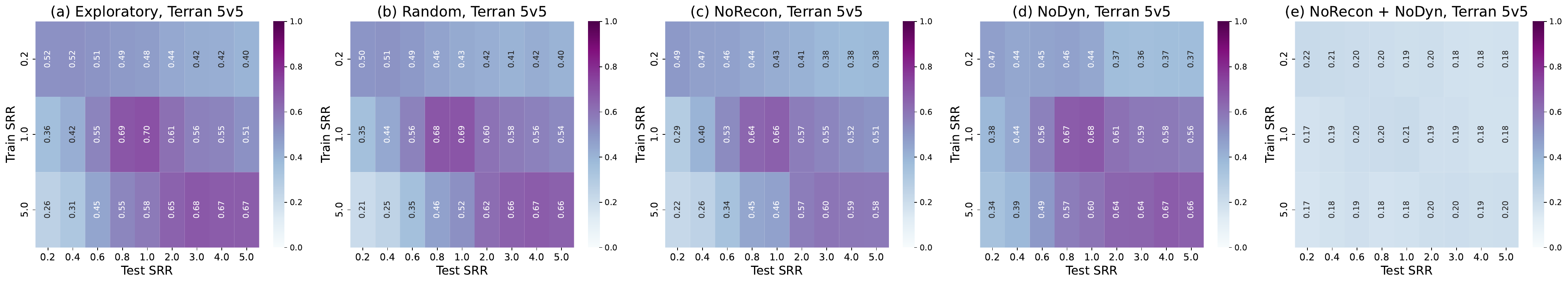}
    \caption{Performance of \algo\ on policy generalizability across various sight ranges under 5 different offline training schemas. The environment setup is the Terran map with a 5v5 agent configuration. (a) Offline training with a \texttt{Exploratory} dataset; (b) Offline training with a \texttt{Random} dataset; (c) Offline training without $L_{recon}$; (d) Offline training without $L_{dyn}$; (e) Offline training without $L_{recon}$ and $L_{dyn}$.}
    \label{fig:ablation}
\end{figure*}
\subsection{Ablation Study}

In this section, we investigate Q3 by conducting ablation studies on the quality of the offline dataset $\mathcal{D}$ for training the communication mechanism and the impact of different loss terms in our objective function~(Eq.~\ref{eq:loss}). For this section, we focus on the Terran map and 5-agent setting.

\noindent\textbf{Data quality.} To examine the influence of data quality of $\mathcal{D}$ on the learned online policy, we train the communication mechanism twice on two types of offline datasets.
\begin{itemize}
    \item \texttt{Exploratory}: Trajectories generated in the exploratory stage of QMIX training, with $6000$ episodes per task.
    \item \texttt{Random}: Trajectories generated from random environment interactions (where actions are chosen uniformly at random for each agent), with $6000$ episodes per task.
\end{itemize}

We then perform online policy learning twice using the two pre-trained communication mechanisms, obtaining two policies. They are evaluated in the same way as in Section~\ref{sec:exp_generalization}. The performance is shown in Figure~\ref{fig:ablation}(a-b). It can be seen that there is a very slight performance decay when you use the \textsc{Random} dataset compared to the \textsc{Exploratory} dataset, indicating that the offline training stage for the communication mechanism can still learn meaningful and informative representations given low-quality data.

\noindent\textbf{Loss terms.} We further examine the influence of the two auxiliary loss objectives, the reconstruction loss $L_{recon}$ and the dynamic loss $L_{dyn}$, by training three communication mechanisms with the weighing factors set to $\alpha=0, \beta=1$, $\alpha=1, \beta=0$, and $\alpha=0, \beta=0$ in Eq.~\ref{eq:loss}, respectively. The online policy learning and evaluation remain the same as the data-quality ablation study. The results are shown in Figure~\ref{fig:ablation}(c-e).

In the first two cases where either the $L_{recon}$ or $L_{dyn}$ is dropped, there are no significant performance decays. Specifically, both terms can help the offline stage learn informative representations individually, while $L_{recon}$ contributes slightly more to the overall performance than $L_{dyn}$. 
However, when both terms are dropped, a significant performance decay is observed as shown in Figure~\ref{fig:ablation}(e). This demonstrates the necessity of adopting at least one of the auxiliary loss objectives, and the combination of the two leads to the most effective representation learning in the offline stage.




%% file: AAMAS2025/sections/conclusion.tex
\section{Conclusion}
In this paper, we introduce \algo, a communication mechanism for improving the generalizability of MARL systems. Specifically, we focus on generalizing the policy to scenarios where the sight range of the agents during execution may not be the same as that during training. 
The communication mechanism in \algo\ is trained offline and is task-agnostic. Utilizing contrastive loss allows agents to effectively align the integration of local observation and incoming messages with the egocentric state, leading to better situational awareness and improved coordination. 

We show that standard benchmark MARL techniques with and without inter-agent communication generalize poorly when the sight range changes. In contrast, we show \algo\ generalizes better across sight ranges. Notably, we find that even random exploration trajectories can be leveraged to learn an effective communication strategy. Our findings suggest that \algo\ provides a robust framework to enhance inter-agent communication, leading towards more adaptive MARL systems.
